# Towards the Distribution of a Class of Polycrystalline Materials with an Equilibrium Defect Structure by Grain Diameters: Temperature Behavior of the Yield Strength

A.A. Reshetnyak[1,2,a)] and V.V. Shamshutdinova[2,b)]

[1]*Center of Theoretical Physics, Tomsk State Pedagogical University, 75, Ave. Comsomolskii, Tomsk, 634061, Russia*
[2]*National Research Tomsk Polytechnic University, 10, Ave. Lenin, Tomsk, 634011, Russia*

[a)] Corresponding author: reshet@tspu.edu.ru
[b)] shamshut@tpu.ru

**Abstract.** We modify a theory of flow stress introduced in [6,7,8] for a class of polycrystalline materials with equilibrium and quasy-equilibrium defect structure under quasi-static plastic deformations. We suggest Maxwell-like distribution law for defects (within dislocation-disclination mechanism) in the grains of polycrystalline samples with respect to grain's diameter. Polycrystalline aggregates are considered within single- and two-phase models that correspond to the presence of crystalline and grain-boundary (porous) phases. The scalar dislocation density is derived. Analytic and graphic forms of the generalized Hall–Petch relations for yield strength are produced for single-mode samples with BCC ($\alpha$-Fe), FCC (Cu, Al, Ni) and HCP ($\alpha$-Ti, Zr) crystal lattices at $T = 300\ K$ with different values of the grain-boundary phase. We derived new form of the temperature-dimensional effect. The values of extremal grain and maximum of yield strength are decreased with raising the temperature in accordance with experiments up to NC region.

## INTRODUCTION

The basic directions in materials science are the search and the control of the internal defect substructure of grains in order to produce the best plastic and strength properties of polycrystalline (PC) aggregates.

An improvement of these properties uses the new technologies, including the vapor deposition method, the ones of severe plastic deformation (SPD), its combination with recrystallization annealing, so on [1,2]. These methods permit for large variability in linear size *d* and orientation of the elements of material microstructure: from nanocrystalline (NC, <100 nm), submicrocrystalline (SMC, 100−500 nm), ultrafine-grained (UFG, 0.5−2 μm), to fine-grained (FG, 2−10 μm), coarse-grained and up to mesopolycrystalline (CG, 10−1000 μm) samples. Experimental research of the physical and-mechanical properties of PC materials (strain hardening coefficient θ, ultimate stress $\sigma_S$, microhardness *H* and also yield strength $\sigma_y$) has shown the features of the hardening mechanism in the transition to UFG, SMC and NC states, marked in the deviation from the empirical Hall–Petch (HP) relation, [3]: $\sigma_y(d) = \sigma_0 + kd^{-1/2}$, (*k* and $\sigma_0$ are respectively the HP coefficient and the frictional stress in dislocations in case of its moving inside the grains). This study has been developed in many works of A.W. Thompson, G. Langford, S.A. Firstov, J.G. Sevillano, R. Armstrong, U.F. Kocks, B.A. Movchan, H. Conrad, V.I. Trefilov, V.V. Rybin, E.V. Kozlov, V.A. Likhachev, Yu.Ya. Podrezov, V.E. Panin, R.Z. Valiev, N.A. Koneva and many others. Up to now there was no satisfactory theoretical justification of the HP law, explanation for the presence and behavior of maxima $\sigma_S, \sigma_y$ and analytical representation for $\sigma = \sigma(\varepsilon, d)$ relation under plastic deformation (PD) beyond the continuum media mechanics concepts, G. Malygin dislocation kinetic model [4,5]. In our papers [6, 7, 8] it was proposed a theoretical model of mechanical energy distribution in each grain of a single-mode isotropic PC material with respect to quantized quasi-stationary zones (levels of a finite width in the energy spectrum of crystal lattice (CL)) under quasi-static PD. In the first approximation the energy spectrum of each grain was chosen as the equal-distant one with step $\frac{1}{2}Gb^3$ (realizing the most probable assembly of dislocations with a unique Burgers vector $b$ and shear modulus $G$) initiating from the zero-level energy $E_0$ of a crystallite without defects up to the level $E_N$ with the maximal number of atoms $N(d) = [d/b]$ in a (full) dislocation located on a dislocation axis. The expressions for statistically determined *scalar density of dislocations* (SDD) $\rho(b_\varepsilon, d, T, \varepsilon)$, *flow stress* (FS), including $\sigma_y$, for single-mode PC under quasi-static PD with velocities: $\dot\varepsilon \sim 10^{-5} - 10^{-3} \, s^{-1}$ in dependence on $d$, for grains, as well as on a value of grain-boundary (weak) unhardening phase were calculated in [7], [8]. It was based on the Bose-Einstein distribution of defects in the grains by its average diameter $d$: $\left(e^{M(\varepsilon)b/d} - 1\right)^{-1}$, for energy scale $M(\varepsilon) = Gb_\varepsilon^3/2k_BT$ (with $M(\varepsilon) \sim 10^2$, when $T \sim 300\,K$, to be inverse of the speed sensitivity), for effective Burgers vector $b_\varepsilon = b(1+\varepsilon)$. However, due to significant distortion of CL in the grains under the acts of defect creation (annihilation), asymmetry in grains orientation, as well as in not polyhedral identity of grains shape such quantum-like distribution (for regular distributed similar grains) for defects may pass in classical Maxwell-Boltzmann distribution, $e^{-M(\varepsilon)b/d}$, of defects with respect to inverse square root of diameter $d^{-1/2}$ in approximation of single dislocation ensemble

The purpose of the paper is to derive the SDD, FS, in particular, yield strength $\sigma_y$, within modified distribution laws of defects in the grains of a PC material.

The paper organized as follows. In the second section we establish the Maxwell-Boltzmann distribution of defects in the grain by its diameters, modify original [6,7] Bose-Einstein distribution and derive the analytical forms for SDD, FS for class of PC materials both in one-phase and in two-phase (with GB) models. In the third section we investigate a new forms of HP law for single-mode samples with HCP ($\alpha$-Ti, Zr) FCC (Cu, Al, Ni) and BCC ($\alpha$-Fe) crystal lattices at $T = 300\,K$. We conclude the paper by final remarks.

## SCALAR DISLOCATION DENSITY, FLOW STRESS IN ONE- AND TWO-PHASE APPROXIMATIONS

Following to [6] we will call the one-phase PC sample by *equilibrium PC sample* (or by one with *equilibrium defect structure* (EDS) if in it after series from $N_0 = N_0(\alpha; T)$ identical combinations (denoted by the letter $\alpha$) of mechanical tests at SPD for the temperature $T$ for all tests with finite integers $N$: $N \geq N_0$, it is established the DS with SDD $\rho = \rho(b_1, \dots, b_k, d, T; N)$: $|\rho(N) - \rho(N_0)| \ll \rho(N_0)$ in the grains of PC sample for each value of average grain's size $d(N)$.

We stress that in the real mechanical tests at SPD the constant value of temperature $T$ is conserved in the volume of the sample only in average, also the *single-mode* property is realized only locally in the PC sample (e.g. obtained under torsion on the Bridgemen's anvils). Such separate equilibrium parts of the PC sample can be mechanically extracted.

## STATIONARY EQUILIBRIUM SCALAR DISLOCATION DENSITY

Let's deduce the stationary ESDD $\rho = \rho(b_1, \dots, b_k, d, T; N)$ in such equilibrium PC sample in approximation of single dislocation ensemble (for least $b_1 = b$) realization $\rho(b, d, T; N) = \rho(b, d, T)$ for all sizes $d$.

The probability for any of possible defects origin within SPD series to occur as we examine the state of EDS of a grain for given $T$ within approximation of the equidistant energy spectrum with a step to be equal to the energy of a unit dislocation in accordance with the Boltzmann distribution (for $\varepsilon = 0$):

$$E_{n+1,n} = E_{n+1}^d - E_n^d = (1/2)Gb^3, \quad n = 0,1,\dots,N = [d/b], \tag{1}$$

$$P(E_n) \equiv Z^{-1} e^{-\frac{\left(\frac{1}{2}\right)Gb^3}{k_BT}\frac{E_n}{E_N}} = Z^{-1} e^{-\frac{\left(\frac{n}{2}\right)Gb^3}{Nk_BT}}, \quad Z = \sum_{k=0}^{N} e^{-\frac{(k/2)Gb^3}{Nk_BT}}. \tag{2}$$

The number $\langle n_d \rangle$ of atoms in dislocation axis (segments) and its average energy $\langle E_d \rangle$ are calculated following the rule of ensemble averaging, according to (2),

$$\langle E_d \rangle = Z^{-1} \sum_{n=0}^{N} E_n e^{-M\frac{E_n}{E_N}} = \frac{1}{2} Gb^3 \left(e^{\frac{Mb}{d}} - 1\right)^{-1}, \quad M = \frac{\frac{1}{2}Gb^3}{k_BT}, \tag{3}$$

$$\langle n_d \rangle = Z^{-1} \sum_{n=0}^{N} n e^{-M \frac{E_n}{E_N}} = \left(e^{Mb/d} - 1\right)^{-1} \equiv f_N(b,d,T). \tag{4}$$

The quantity $\langle n_d \rangle$ coincides with the Bose-Einstein distribution function $f_N^{BE}(b,d,T)$ for the occurrence of a dislocation of energy $\langle E_d \rangle$ in an arbitrary grain of the PC sample with EDS for fixed $T$. From our arguments above on significant distortion of CL in the grains under the acts of defect creation (annihilation), asymmetry in grains orientation and not polyhedral shape of grains the function $f_N(b,d,T)$ may pass in $f_N^{MB}(b,d,T) \sim e^{-Mb/d} = e^{-\frac{Gb^3}{2k_BT}b/d}$.

To determine Maxwell-Boltzmann distribution of number of defects per volume of equilibrium PC sample, $\bar{n} = \bar{N}/V$, (defect concentration) in inverse square root of average grain diameter $d^{-1/2}$ we look for $d\bar{n}$ following the standard Maxwell arguments in view of isotropic grain orientation in the form

$$\frac{d\bar{n}}{\bar{n}} = 4\pi \left(\frac{Mb}{\pi}\right)^{3/2} x^2 e^{-Mbx^2} dx = f_N^{MB}(b,x^{-2},T)dx, \quad x = d^{-1/2}. \tag{5}$$

The function $f_N^{MB}(b,x^{-2},T)$ obeys standard normalization property: $\int_0^\infty f_N^{MB} dx = 1$ for a density of probability function of defect distribution in inverse square root of grain diameter, so that $d\bar{n}$ is the number of defects in unique PC volume from the interval $[d^{-1/2}, d^{-1/2} + d(d^{-1/2})]$ for grain diameters.

The equilibrium SDD $\rho = \rho(b,d,T)$ under SPD at the crystalline phase (far from GB) of the single-mode PC sample with average diameter $d$ (with, $L_\Sigma$ being the sum of the lengths of all dislocations) can be presented as follows

$$\rho = \frac{L_\Sigma}{V} = b\bar{n} f_N^{MB}(b,d,T) b^{-1/2} = 4\pi\bar{n} \left(\frac{M}{\pi}\right)^{3/2} \frac{b^2}{d} e^{-M\frac{b}{d}}. \tag{6}$$

The SDD $\rho(b,d,T)$ reaches the maximum $\rho_{max}(b,d_{0m},T)$ at extremal grain size $d_{0m}$ which is found from the condition, $\partial \rho / \partial d = 0$

$$d_{0m} = Mb = \frac{Gb^4}{2k_BT} \quad (of\ order\ 10^{-8}\text{-}10^{-7} m), \quad \rho_{max} = \frac{4}{\sqrt{\pi}} \bar{n} M^{1/2} b e^{-1}, \quad (\rho_{max} \sim \bar{n} T^{-1/2}). \tag{7}$$

For an average grain we find $\bar{n}$ from the CG limit of $\rho(b,d,T)$ (when $e^{-Mb/d} \to 1$) according to H.Conrad [2, 9], but depending on temperature: through a $d$-independent factor $A(T)$

$$\lim_{d \gg b} \rho(b,d,T) \sim A(T)/(bd) \quad \Rightarrow \quad \bar{n}|_{d \gg b} = \langle m_0 \rangle M^{-p/2} b^{-3}, 0 < p < 1, \tag{8}$$

where $\langle m_0 \rangle$ is a fitting parameter (like to a polyhedral parameter [6]), which for $T = 300\,K$ should related to Conrad constant $\langle \varepsilon \rangle$, as $\langle \varepsilon \rangle = \frac{4}{\sqrt{\pi}} \langle m_0 \rangle M^{(3-p)/2}$, (for$\langle \varepsilon \rangle = \max_i \ln(c_i/d_i)$, $i = x,y,z$ with $d_i(c_i)$ being the sizes of the grain before (after) SPD applied to PC sample). The indicator $p$ depends on type of PC material. Its range provides increasing of defect concentration and decreasing of SDD maximum (reflecting dependence on GB input) with the growth of temperature: $\rho_{max}(T_1) < \rho_{max}(T_2)$ for $T_1 > T_2$. As the result, we get

$$\rho = \frac{4}{\sqrt{\pi}} \langle m_0 \rangle M^{(3-p)/2} \frac{1}{bd} e^{-M\frac{b}{d}}. \tag{9}$$

A modification of Bose-Einstein distribution function $f_N^{BE}(b,d,T)$ may be immediately realized following to (5):

$$\frac{d\bar{n}}{\bar{n}} = \frac{4(Mb)^{3/2}}{\sqrt{\pi}\zeta(3/2)} x^2 \left(e^{Mbx^2} - 1\right)^{-1} dx = f_N^{BE}(b,x^{-2},T) dx \tag{10}$$

for $x = d^{-1/2}$ with using the improper integral: $\int_0^\infty x^2 \left(e^{Mbx^2} - 1\right)^{-1} dx = (\sqrt{\pi}/4)\zeta(3/2)(Mb)^{-3/2}$, with Riemann $\zeta$-function, $\zeta(3/2) = 2.61238$, that is led to the respective form of SDD

$$\rho^{BE} = b\bar{n} f_N^{BE}(b,d,T) b^{-1/2} = \frac{4}{\sqrt{\pi}\zeta(3/2)} \bar{n} M^{3/2} \frac{b^2}{d} \left(e^{Mb/d} - 1\right)^{-1} \tag{11}$$

with account of Conrad limit (8) $\bar{n}|_{d\gg b} = \langle m'_0 \rangle M^{-p'/2} b^{-2} d^{-1}$: for $0 < p' < 1$. The maximum $\rho_{max}^{BE}(b,c,T)$ at extremal grain size $d_0$ is determined as in (7) from the equation $\partial \rho^{BE}/\partial d = 0$ ($\Rightarrow [e^y - 1 - 0{,}5ye^y] = 0$) [6]

$$d_0 = \frac{1}{1.59362} Mb = \frac{1}{1.59362} d_{0m}, \quad \rho_{max}^{BE} = 4 \frac{1.59362}{\sqrt{\pi}\zeta(3/2)} \bar{n} M^{1/2} b (e^{1.59362} - 1)^{-1} \sim \bar{n} T^{-1/2}. \tag{12}$$

For the indicators range $0 < p, p' < 1$ the CG concentration of defects increased with growth of $T$. As the result, we come to

$$\rho^{BE} = \frac{4}{\sqrt{\pi}\zeta(3/2)} \langle m'_0 \rangle M^{(3-p')/2} \frac{1}{d^2} (e^{Mb/d} - 1)^{-1}. \tag{13}$$

## ε-EVOLUTION FOR THE EQUILIBRIUM SCALAR DISLOCATION DENSITY AND FLOW STRESS

Our next aim is to find the equilibrium SDD under quasi-static loading (by tension). To this end we determine $t$- (*equally $\varepsilon$-*) *evolution* of EDS naturally suppose that probabilities $P(E_n, \varepsilon)$ (2) for any of possible defects origin with preservation of $\varepsilon$-modified equidistant energy spectrum (1) satisfy to the linear differential equations with initial conditions $P(E_n, 0) = P(E_n)$

$$\frac{d}{d\varepsilon} P(E_n, \varepsilon) = f_n(\varepsilon) P(E_n, 0) \quad \Leftrightarrow \quad \frac{d}{dt} P(E_n, \dot{\varepsilon}t) = \dot{\varepsilon} f_n(\dot{\varepsilon}t) P(E_n, 0), \quad n = 1, 2, \ldots, N. \tag{14}$$

We choose as the solution for the equations (14) the Boltzmann-like distribution

$$P(E_n, \varepsilon) \equiv Z^{-1}(\varepsilon) e^{-\frac{(1/2)Gb_\varepsilon^3 E_n}{k_B T E_N}} = Z^{-1}(\varepsilon) e^{-\frac{(n/2)Gb_\varepsilon^3}{Nk_B T}}, \quad Z(\varepsilon) = \sum_{k=1}^{N} e^{-\frac{(k/2)Gb_\varepsilon^3}{Nk_B T}} \tag{15}$$

with *effective Burgers vector* $b_\varepsilon = b(1 + \varepsilon)$ and also the *effective dislocation energy* $E_n(\varepsilon) = (n/2)Gb_\varepsilon^3$.

Augmenting the arguments when deriving stationary SDD by counting the number of dislocations that appeared during the time $t = \varepsilon/\dot{\varepsilon}$ we get for the case of Maxwell-Boltzmann distribution (beyond the conditional elastic limit $\varepsilon > \varepsilon_{0.05}$) for ε-dependent part of total SDD $\rho_\varepsilon = \rho(b, d, T; \varepsilon)$

$$\rho_\varepsilon - \rho = \frac{4}{\sqrt{\pi}} m_0 \varepsilon M(\varepsilon)^{(3-p)/2} \frac{1}{bd} e^{-M(\varepsilon)\frac{b}{d}} + o(\varepsilon). \tag{16}$$

Under assumption of a validity of Taylor [10] deformation (dislocation) hardening law in whole range of grain diameter $d$, including NC, SMC and UFG materials, we have (following the results of [6] due to the interaction of dislocations) for tangential FS $\tau \sim Gbl^{-1} \sim Gb\rho^{1/2}$ and also for $\varepsilon > \varepsilon_{0.2} = 0{,}002$ that

$$\tau(\varepsilon) = \tau_f + \alpha Gb(\rho_\varepsilon - \rho)^{1/2}. \tag{17}$$

Here $\tau_f$ and $\alpha$ are respectively a frictional stress at the interaction of propagating dislocations with lattice defects and obstacles of non-deformation origin a dislocation and an interaction constant, varying for different materials in the range (0.1–0.4),. Taking into account that the FS of a PC sample $\sigma(\varepsilon)$ is proportional to $\tau(\varepsilon)$: $\sigma(\varepsilon) = m\tau(\varepsilon)$, (with Taylor factor $m = 3.05$) according to (16), (17), we obtain for isotropic GB distribution for

$$\sigma(\varepsilon) = \sigma_0(\varepsilon) + \alpha mG \left(\frac{b}{d}\right)^{1/2} \left(\frac{4}{\sqrt{\pi}} m_0 \varepsilon M(\varepsilon)^{(3-p)/2}\right)^{1/2} e^{-\frac{M(\varepsilon)b}{2d}} + o(\varepsilon), \quad \sigma_0(\varepsilon) = m\tau_f(\varepsilon). \tag{18}$$

Expression (18) is – with accuracy up to higher orders in $\varepsilon$ – by the basic analytical result of applying our statistical model to the determination of FS at the crystalline phase of a single-mode equilibrium PC sample with isotropic GB orientation in all grain ranges under QS loading (within single dislocation ensemble approximation).

In case of multiple $k$ dislocation ensembles due to probabilities the $\varepsilon$-dependent part of equilibrium SDD $\rho_\varepsilon - \rho = \sum_j (\rho_\varepsilon - \rho)_j$ (and therefore FS) should have a more complicated form:

$$\rho_\varepsilon - \rho \sim m_0 \varepsilon \sum_j M_j(\varepsilon)^{(3-k)/2} \frac{1}{b_j d} e^{-M_j(\varepsilon)\frac{b_j}{d}} + o(\varepsilon),$$

and presents a separate research.

For general form of GB disorientation ($\alpha_i(\varepsilon)$, $N_i$), the integral FS $\sigma_\Sigma = \sigma(\varepsilon, \{\alpha_i\})$ for PC sample with quasi-equilibrium defect structure may be determined additively according to

$$\sigma_\Sigma = \sigma_0(\varepsilon, \{\alpha_i\}) + \alpha m G \left(\frac{b}{d}\right)^{1/2} \sum_i N_i K(\alpha_i) \left(\frac{4}{\sqrt{\pi}} m_0 \varepsilon M(\varepsilon_i)^{(3-p)/2}\right)^{1/2} e^{-\frac{M(\varepsilon)_i b}{2d}} + o(\varepsilon) = \qquad (19)$$
$$= \sigma_0 + \sum_i N_i K(\alpha_i) \sigma_i(\varepsilon_i) + o(\varepsilon_i) \quad \text{for} \quad \varepsilon = \sum_i N_i \varepsilon_i, \quad \sum_i N_i = 1,$$

where nonnegative numbers $K(\alpha_i)$ ($K(\alpha_i) \leq 1$) determine the value of GB disorientation (from EBSD analysis, see e.g. [11]) in respective $N_i$ part of PC sample with strength $\sigma_i$ and strain $\varepsilon_i$. Here, we use a division the sample on parts (not, in general, connected) with subvolumes $V_i$, $i = 1,2,\ldots,l$, $\sum_i V_i = V$, where the average angle $0 \leq \alpha_i < \pi/2$ of GB disorientation is restricted within small conus ($\alpha_i$, $\alpha_i + \Delta\alpha_i$) subject to the ordering, $\alpha_i < \alpha_{i+1}$. Such division may be determined by means of EBSD (also by SEM, energy dispersive X-ray) analysis (see, Fig. 3 in [6]) applied to the real PC sample with respect to GB angle and grain size distributions.

We suppose the results (18)-(19) are applicable for quasi-equilibrium PC samples not only for yield strength but as well for the stages of parabolic and linear hardening down to the stages of pre-fracture according to the Backofen-Considére condition. Notice, that the values $\sigma_0(0) = \sigma(0) = 0$ and $\sigma_0(0.002) = \sigma_0$ are given for $\varepsilon = 0.002$ in (18)-(19). Dependence $\sigma(\varepsilon)$ (18) determines the FS maximum $\sigma_m(\varepsilon) = \sigma(\varepsilon)|_{d=d_0}$ in dependence on the extremal grain size $d_{0m}(\varepsilon, T) = M(\varepsilon)b = d_{0m}(1 + \varepsilon)^3$ with $d_{0m}$ defined in (7).

For completeness, we derive an expression for modified FS of Bose-Einstein defect distribution at the crystalline phase of a single-mode equilibrium PC sample with isotropic GB distribution obtained according to (13), (17)

$$\sigma^{BE}(\varepsilon) = \sigma_0(\varepsilon) + \alpha m G \frac{b}{d} \left(\frac{4}{\sqrt{\pi}\zeta(3/2)} m'_0 \varepsilon M(\varepsilon)^{(3-p')/2}\right)^{1/2} \left(e^{M(\varepsilon)b/d} - 1\right)^{-1/2} + o(\varepsilon). \qquad (20)$$

Note, that we have inverse *temperature-dimension effect* (for $p < 1$) as compared to one proposed in [6,7]. Namely, for a maximum of SDD we have both for Maxwell-Boltzmann and for Bose-Einstein distributions $\sigma_m(T) \sim T^{(p-1)/4}$ and, hence, $\sigma_m(T_1) < \sigma_m(T_2)$ for $T_1 > T_2$. For $p > 1$ we have in NC region the growth of $\sigma_m(T)$ with growth of $T$, i.e. revealing direct [6,7] *temperature-dimension effect* that should be tested.

## HALL–PETCH LAW FOR $\alpha$-TI, ZR, CU, AL, NI, $\alpha$-FE, IN ONE-PHASE AND TWO-PHASE APPROXIMATIONS

For equilibrium CG materials, the normal HP law for yield strength when $\varepsilon = 0.002$ implies a relation between HP coefficient $k(\varepsilon)|_{\varepsilon=0.002}$ and polyhedral parameter $m'_0$ denoted as $m'^{CG}_0$ in the CG limit. To preserve a dependence on indicator $p'$ from above CG limit (due to different value of $k(\varepsilon)|_{\varepsilon=0.002}$ for the same material) we present it as $(p' + \Delta p - \Delta p)$ and will establish relation of HP coefficient with $\sigma^{BE}(\varepsilon)|_{d \gg b}$ for $p' + \Delta p$:

$$\sigma^{BE}(\varepsilon)|_{d \gg b} = \sigma_0(\varepsilon) + k(\varepsilon) d^{-1/2}, \quad k(\varepsilon) = 2\alpha m G \left[m'^{CG}_0 \frac{\varepsilon b (M(\varepsilon))^{(1-p'-\Delta p)/2}}{\sqrt{\pi}\zeta(3/2)}\right]^{1/2}\bigg|_{\varepsilon=0.002} \qquad (21)$$
$$\Rightarrow \alpha^2 m'^{CG}_0 = \frac{\sqrt{\pi}\zeta(3/2)}{4\varepsilon b} \frac{k^2(\varepsilon)}{(mG)^2} (M(\varepsilon))^{(p'+\Delta p-1)/2}\bigg|_{\varepsilon=0.002} \Leftrightarrow m'^{CG}_0 = \frac{\sqrt{\pi}\zeta(3/2)}{4\varepsilon b} \frac{k^2(\varepsilon)}{(\alpha m G)^2} (M(\varepsilon))^{(p'+\Delta p-1)/2}\bigg|_{\varepsilon=0.002}.$$

The correspondence above permits to establish explicit interrelation among the empirical and theoretical HP relations for number of materials and to discover a *temperature–dimension effect* (see as well [2,6-8]).

The maximum undergoes a shift to the region of larger grains for decreasing temperatures. A coincidence is well established between the theoretical and experimental data on $\sigma_y$ for the materials with EDS for HCP ($\alpha$-Ti, Zr), FCC (Cu, Ni, Al) and BCC ($\alpha$-Fe), crystal lattices with closely packed grains at $T = 300\ K$.

**TABLE 1.** The values of $\sigma_0$, $\Delta\sigma_m^{BE} = (\sigma_m^{BE} - \sigma_0)$, $E_d^{L_e}$, $k$, $m'_0^{CG}$, $\alpha$ in HCP, FCC and BCC PC metal samples with $d_0$, $d_{0m}$, $b$, $G$ and indicator $p' + \Delta p = p = 0.5$ from (7), (12), (20), (21) and from [2,4,7]

| CL | BCC | FCC | | | HCP | |
|---|---|---|---|---|---|---|
| **Material** | $\alpha$-Fe | Cu | Al | Ni | $\alpha$-Ti | Zr |
| $\sigma_0$, MPa | 170 (anneal.) | 70 (anneal.); 380 (cold-worked) | 22 (anneal. 99.95%); 30 (99,5%) | 80 (anneal.) | 100 (~100%); 300 (99.6%) | 80-115 |
| $b$, nm | $\frac{\sqrt{3}}{2}a = 0.248$ | $\frac{a}{\sqrt{2}} = 0.256$ | $\frac{a}{\sqrt{2}} = 0.286$ | $\frac{a}{\sqrt{2}} = 0.249$ | $a = 0.295$ | $a = 0.323$ |
| $G$, GPa | 82.5 | 44 | 26.5 | 76 | 41.4 | 34 |
| $k$, MPa·m$^{1/2}$ | 0.55-0.65 $10^{-5}-10^{-3}$m | 0.25 $10^{-4}-10^{-3}$m | 0.15 $10^{-4}-10^{-3}$m | 0.28 $10^{-5}-10^{-3}$m | 0.38-0.43 $10^{-5}$-$10^{-3}$m | 0.26 $10^{-5}-10^{-3}$m |
| $\alpha$ | – | 0.38 | – | 0.35 | 0.97 | – |
| $E_d^{L_e} = Gb^3/2$, eV | 3.93 | 2.31 | 1.96 | 3.72 | 3.33 | 3.57 |
| $\alpha^2 m'_0^{CG}$ | 3.18-4.43 | 2.55 | 2.36 | 0.98 | 5.27-6.75 | 3.29 |
| $d_0$, nm | 23.8 | 14.4 | 13.5 | 22.3 | 23.9 | 28.2 |
| $d_{0m}$, nm | 37.9 | 22.9 | 21.5 | 35.5 | 38.1 | 44.9 |
| $\Delta\sigma_m^{BE}$, GPa | 2.27-2.69 | 1.33 | 0.83 | 1.19 | 1.57-1.77 | 0.99 |
| $\alpha^2 m_0^{CG}$ | 0.008-0.011 | 0.011 | 0.012 | 0.003 | 0.016-0.020 | 0.009 |

The values of $k$ at $\varepsilon = 0.002$ are taken, e.g., for $\alpha$-Fe, Cu, Ni [6,7], Al [10], Zr, $\alpha$-Ti [2] within the grain range enclosed in the frames.

**TABLE 2.** Most probable crystallographic sliding systems at $T = 300$ K for $\alpha$-Fe, Cu, Al, Ni, $\alpha$-Ti, Zr for BCC, FCC (Miller indices), HCP lattices (Miller–Bravais indices)

| | Ni | Al | Cu | $\alpha$-Fe | $\alpha$-Ti | Zr |
|---|---|---|---|---|---|---|
| **Plane** | {111} | {100},{{111} | 111} | {110},{112},{123} | $(10\bar{1}0),(10\bar{1}1)$ | $(10\bar{1}0)$ |
| **Direction** | <110> | <110> | <110> | <111> | $[2\bar{1}\bar{1}0]$ | $[11\bar{2}0]$ |

The graphic dependences $\sigma_y = \sigma_y(d^{-1/2})$ for the hard (crystalline) phase of PC aggregates of $\alpha$-Ti, Zr, Cu, Al, Ni, $\alpha$-Fe with closely-packed randomly oriented grains, to be homogeneous with respect to their size (single-mode case) at $T = 300\ K$, are presented on the Fig. 1.

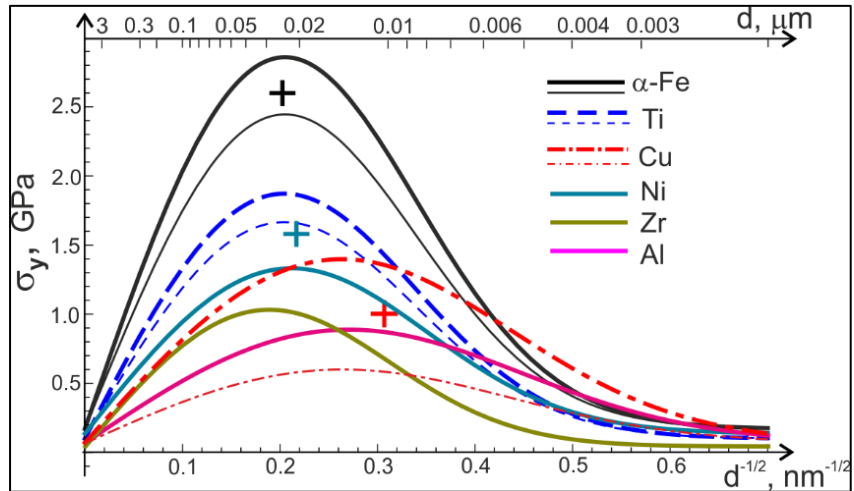

**FIGURE 1.** Graphic dependence for HP law with an additional upper scale, to be changing within range $(\infty;0)$ with the inverse quadratic scale, with size of grains $d$ given in μm. For the values of both axes the correspondence (100; 1.6; 0.1; 0.044; 0.025; 0.011; 0.006; 0.004; 0.003) μm ↔ (0.005; 0.015; 0.1; 0.15; 0.2; 0.3; 0.41; 0.5; 0.57) nm$^{-1/2}$ is

true for the respective values on the lower axis. By the colored cross for $\alpha$-Fe (black), Cu (red), Ni (green) it is shown the experimental maxima [2], [12], [13]. Here, the thin black for $\alpha$-Fe, dashed blue for $\alpha$-Ti and dot-dashed red for Cu lines are derived for choices of k($\alpha$-Fe)=0,55; k($\alpha$-Ti)=0,38; k($Cu$)=0,15 respectively,

## CONCLUSION

We proposed both the new form of Maxwell-Boltzmann distribution (5) (as compared to idea of [6]) and modified Bose-Einstein distribution (10) of number of defects (dislocations) per volume of equilibrium PC sample in inverse square root of average grain diameter. The Maxwell-Boltzmann distribution may follow from a significant distortion of CL in the grains under the acts of defect creation (annihilation), asymmetry in grains orientation and not polyhedral identity of grains shape. We deduced the stationary scalar dislocation density (9) for the Maxwell-Boltzmann and one given by (13) for Bose-Einstein distributions, which have the correct coarse-grained Conrad's limit (8) and satisfactory temperature dependence, providing decreasing both the extremal grain size and maximum of SDD (7), (13) with growth the temperature within the same phase of PC materials. We suggested a procedure of $\varepsilon$-evolution for the equilibrium scalar dislocation density (16) under quasi-static loading and analytically derived the generalized law of flow stress $\sigma(\varepsilon)$ for Maxwell-Boltzmann (18) and of $\sigma^{BE}(\varepsilon)$ (20) for Bose-Einstein defect distributions for single-mode PC materials with isotropic GB disorientation. Both models depend on two fitting parameters: polyhedral parameter $m_0$ and indicator $p$ ($m'_0$, $p'$ for Bose-Einstein case in (20)) whose concrete values should follow from tested CG limit for Hall-Petch relation and temperature behavior of flow stress. We generalized the dependences for FS for general form of GB disorientation ($\alpha_i(\varepsilon)$, $N_i$) with different ordered parts of PC sample with weights $N_i$ characterized by GB angles $\alpha_i(\varepsilon)$ and derive the integral FS $\sigma_\Sigma = \sigma(\varepsilon, \{\alpha_i\})$ (19) for PC sample with quasi-equilibrium defect structure with second (GB) phase.

From Tables 1, 2 and Figure 1, the data for the Hall–Petch law at $T = 300\,K$ in homogeneous single-mode PC samples of $\alpha$-Ti, Zr, Cu, Al, Ni, $\alpha$-Fe for the one-phase model with equilibrium defect structure are constructed for modified Bose-Einstein defect distribution and it very well correspond to experimental data [2, 12, 13]. Both one-phase and two-phase model provide usual temperature-dimension behavior (see, e.g. [7]) in whole range of grains of single-mode PC aggregates. We stress, that the Bose-Einstein-like distribution [4] for the defects in equilibrium PC samples was corrected to realize usual temperature-dimension effect.

The proposed theoretical model implies interesting perspectives for application for new PC composite materials in cosmic industry, aircraft and has been experimentally tested with use of ultrafine-grained PC and BT1-0 $\alpha$-Ti alloy samples [15]**.** The model permits one to find an analytic form for the $\sigma$-$\varepsilon$ dependence of PC materials in the whole range of grain sizes, values of temperature and accumulated strain, correctly reflecting the experimental data. Further developments are expected for PC samples both with multiple dislocation ensembles and with various dispersion phases, It allows a natural inclusion of twinning defects into the Taylor dislocation hardening mechanism due to their appearance as combinations of different (also partial) dislocations, especially in the NC segment.

## ACKNOWLEDGMENTS

The authors are grateful to the research fellows of ISPMS SB RAS for valuable discussions. The paper was carried out as part of an initiative project of Tomsk State Pedagogical University and Tomsk Polytechnic University.